\documentclass[a4paper,fleqn]{cas-dc}

\usepackage{graphicx}
\usepackage{siunitx,bm,amsmath}
\usepackage[export]{adjustbox}
\usepackage{multirow}
\usepackage[authoryear]{natbib}

\def\tsc#1{\csdef{#1}{\textsc{\lowercase{#1}}\xspace}}
\tsc{WGM}
\tsc{QE}

\definecolor{coral}{rgb}{1.0, 0.5, 0.31}

\begin{document}
\let\WriteBookmarks\relax
\def\floatpagepagefraction{1}
\def\textpagefraction{.001}

\shorttitle{Hybrid physics-data driven spectral forecasts of semisubmersible response}    

\shortauthors{Milne, Astfalck, Zed, Lee-Kopij, Cripps}  

\title [mode = title]{Hybrid physics-data driven spectral forecasts of semisubmersible response} 



%

\author[1,2]{I.A. Milne}

\cormark[1]


\ead{ian.milne@uwa.edu.au}



\affiliation[1]{organization={School of Earth and Oceans, The University of Western Australia},
            city={Crawley},
            postcode={6009}, 
            state={WA},
            country={Australia}}
						
\affiliation[2]{organization={Woodside Energy},
            addressline={11 Mount Street}, 
            city={Perth},
            postcode={6000}, 
            state={WA},
            country={Australia}}

\affiliation[3]{organization={Department of Mathematics and Statistics, The University of Western Australia},
            city={Crawley},
            postcode={6009}, 
            state={WA},
            country={Australia}}

\author[3]{L.A. Astfalck}
\author[2]{M. Zed}
\author[2]{J. Lee-Kopij}
\author[3]{E. Cripps}





\cortext[1]{Corresponding author}



\begin{abstract}
A framework for probabilistic forecasting of vessel motion is developed and validated for a semisubmersible operating in long period swell. Bayesian statistical methods are applied to predictions of the heave response from a physics model using numerical wave spectra and measured motion data. Model diagnoses motivate an additional level of complexity required for the error
structure in the Bayesian model, specifically to account for heteroskedasticity and time-correlated errors. The hybrid model forecasts were evaluated during periods where the heave resonance and cancellation frequencies were excited. The method is demonstrated to be effective for providing reliable quantification of uncertainty and correcting bias in the raw physics model predictions. This justifies its value for improving the efficiency and safety of offshore operations.
\end{abstract}




\maketitle

\section{Introduction}\label{}
Offshore operations have historically relied on forecasts of the seastate for assessing weather related risk. Most commonly these forecasts are of integrated spectral parameters such as the significant wave height ($H_s$), with respect to the time horizon. Many offshore operations involving floating bodies are, however, fundamentally restricted by thresholds associated with wave induced hydrodynamic response. For instance, drilling may be limited by the vertical travel of the drillstring and a heavy-lift by the velocity of the crane tip. The hydrodynamic response to surface waves can vary appreciably between the different hull forms and displacements (i.e. mass) of the vessels employed across the offshore industry. Notably for regions where long period swell is prevalent, the hydrodynamic heave resonance may be linearly excited and result in large motions. Guidance based only on the seastate can therefore compromise efficiency, safety and cost of operations. 

Response-based forecasts can inherently facilitate superior maritime support for offshore operations.
While wave-by-wave forecasts remain challenging, phase-averaged stochastic or spectral-based response forecasts can be readily operationalised. These require knowledge of only the directional wave energy spectra, $S\left(f,\theta\right)$, and the Response Amplitude Operators (RAOs), the latter describing the hydrodynamic response of the particular vessel to the wave amplitude.
Statistics associated with the short term vessel response can subsequently be readily estimated.
Such an approach was recommended by regulation agencies two decades ago \citep[see][]{HSE2005}. Since then studies such as \cite{Milne2016,Milne2018,Milne2025} have demonstrated the benefits for a Floating Production Offshore and Storage (FPSO) and drillships operating in a complex swell dominated seastates.

Apart from ship-shaped vessels, semisubmersibles have been widely utilised as substructures since the 1960s to support offshore drilling, oil and gas production, heavy-lift cranes and more recently, floating wind turbines.  
Semisubmersibles have particularly attractive hydrodynamic motion characteristics; notably the attenuation of vertical motion at shorter wave periods. 
However, they are weakly damped systems and can be linearly excited in heave by long period swell. In contrast to the Gulf of Mexico and North Sea where numerous semis operate, for regions affected by long period swell such as North West Australia and West Africa, the heave resonance can therefore present a criticality.

The near directionality invariant heave response of a typical semisubmersible suggests that formulating a spectral-based forecast of the response would be simpler than for a ship-shaped vessel. In particular, one could work directly with the 1-D wave (frequency) spectra without the need to resolve directionality or spreading. However, despite these simplifications the form of the heave RAO at long wave periods is characterised by a cancellation point and resonance, at which the response amplitudes are seastate dependent \citep{Clauss1992}. Since swells often originate at considerable distance away they are typically not well resolved by either wave models or wave buoys in frequency space and forecasting the wave energy spectra is also inherently challenging. Further compounding the challenge is that in contrast to parameterised forecasts, the wave spectra available for vessel motion prediction are typically sourced as raw outputs from regional numerical models. These factors combined can lead to potentially significant uncertainties in the heave response forecasts.  


This issue is addressed here by updating raw model, physics-based forecasts of heave response at a location where a semisubmersible rig is operating and measured heave motions are available. The spectra induced RAO forecasts are the state-of-the-art, industry deployed methods described in \cite{Milne2018}. The measured heave motions are particularly novel in that excitation at the heave cancellation and resonant frequencies from long period swell is regularly observed and provide strong empirical evidence of the system at the location of interest. 

We demonstrate a hybrid physics-data driven approach, whereby hybrid approach we mean embedding raw physics-based heave motion forecasts within a data-driven Bayesian statistical model which is informed by the measured heave motions. The result is a method that updates and refines the forecast of the physical model in accordance with recent observations and provides a probabilistic measure of our confidence in the updated forecast to inform operational decisions.  Extending the work of \cite{Loake2022}, who consider a hybrid approach to forecasting sea surface winds for short forecast horizons of 0 and 6 hours, we consider the more challenging task of forecasting vessel heave motions, at forecast horizons of between 0 and 96 hours.  We show that the hybrid approach outperforms the stand alone spectra induced RAO forecasts in terms of the proper scoring rules designed for probabilistic forecasts \citep[see,][]{gneiting2007probabilistic,Astfalck2023}, and provide well calibrated probabilistic forecasts for both longer term (24, 72 and 96 hour) and shorter term (0, 6 and 12 hour) time horizons.

The remainder of the paper is set out as follows. Section \ref{sec:phys_approach} outlines the physics-based approach, reviewing the fundamental hydrodynamics associated with heave response of semisubmersibles, 
 and surface wave forecasts. Section \ref{sec:hybrid} describes the Bayesian hybrid forecast approach and reviews proper scoring rules for probabilistic forecasts.  Section 4 evaluates the performance of the model against the measured motions and demonstrates specific approaches to improve forecasts of the heave response at long wave periods. Section 5 contextualizes the study and discusses possible data-driven approaches to reduce inherent uncertainty in both underlying the hydrodynamic response model and forecasts of swell energy, with conclusions presented in Section 6.


\section{Physics-based approach}\label{sec:phys_approach}
In this section, we describe the physics based approach to compute the motion statistics from wave spectra and the vessel RAOs which serve to elucidate various potential sources of uncertainty. We first introduce the preliminary hydrodynamics for submersibles and surface wave forecasting required to discuss the current state of the art vessel and industry employed semisubmersible response forecasts as described in \cite{Milne2018}.  We do so in the context of a semisubmersible operating in the North West Shelf of Australia.

\subsection{Semisubmersible hydrodynamics}
\begin{figure}
    \centering
    \includegraphics[scale=0.8]{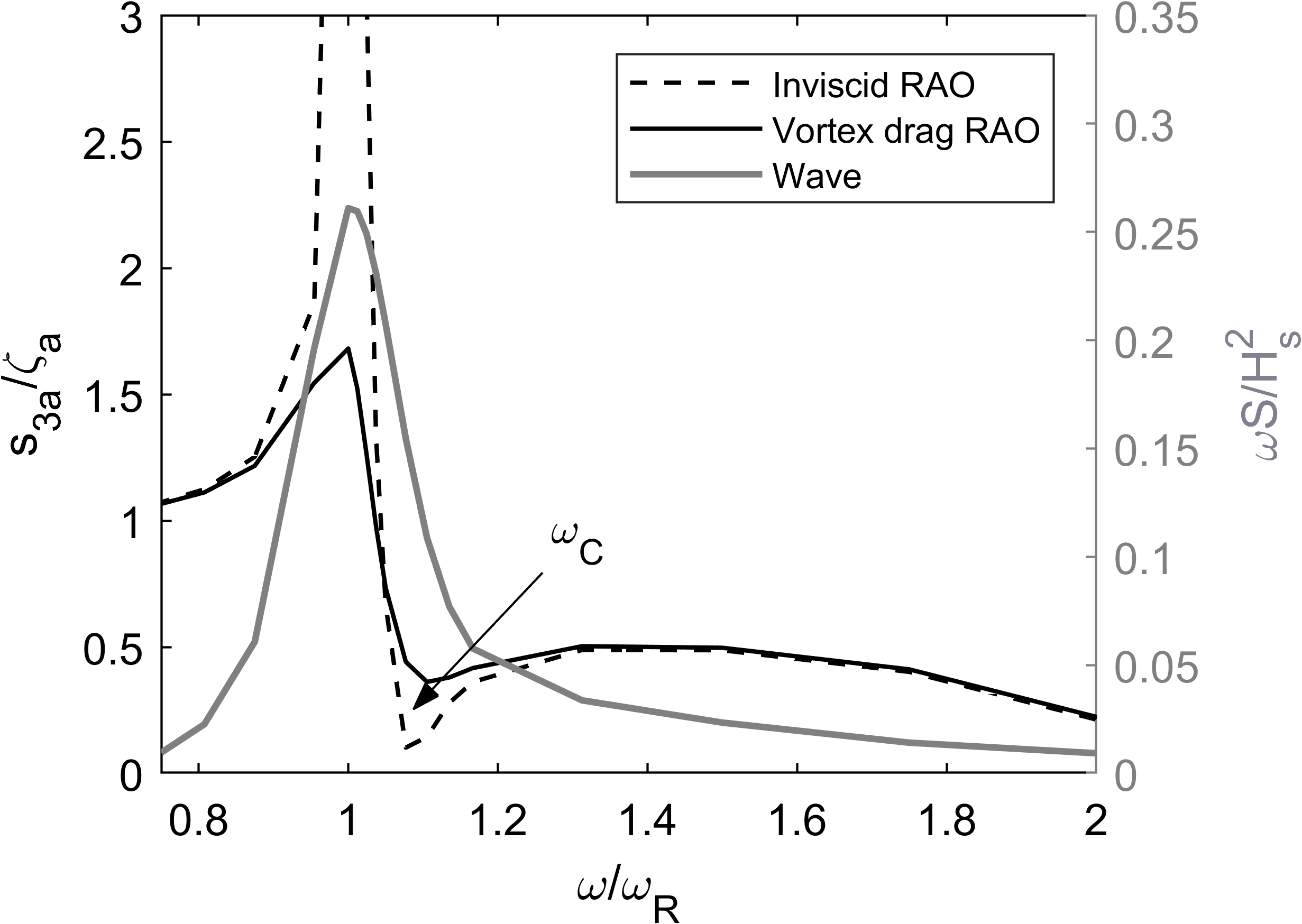}
    \caption{Example of the heave RAO a semisubmersible MODU with and without vortex drag, with wave energy from ocean swell coinciding with the heave resonance and cancellation frequencies.}
    \label{fig:RAOSchematic}
\end{figure}

Conventional semisubmersibles used in offshore drilling and production comprise multiple vertical columns and horizontal pontoons to provide the required stability and buoyancy. Figure~\ref{fig:RAOSchematic} shows the  frequency dependent characteristics of the amplitude of the hydrodynamic heave RAO ($=s_{3a}/{\zeta_a}$, where $\zeta_a$ is the incident wave amplitude associated with frequency component $\omega$) for a typical twin pontoon, four column, semi. This demonstrates how the resonant ($\omega_R$) and cancellation ($\omega_C$) frequencies can be linearly excited by wave energy associated with long period swell. 

On the basis that the hydrodynamic response is inertia-dominated and diffraction effects are comparatively small, the heave motion response may be predicted through application of the Morison equation. Assuming deep water, the magnitude of the heave RAO is given by \cite{Clauss1992} as
\begin{equation}
    \frac{s_{3a}}{\zeta_a} = \frac{F_{3a}}{c_{33}\zeta_a\sqrt{\left[1-\left(\frac{\omega}{\omega_R}\right)^2\right]^2+\left[\frac{b_{33}\omega}{c_{33}}\right]^2}},
\end{equation}
where $b_{33}$ and $c_{33}$ are the damping and restoring coefficients associated with the vertical response. The excitation force, $F_{3a}$, comprises both inviscid and viscous contributions. 
Nonlinearities with respect to wave amplitude in the vertical response are primarily a consequence of vortex shedding around the pontoons. This gives rise to a velocity-dependent drag force which is generally quadratic in nature.

At higher wave frequencies (typically corresponding to wave periods less than approximately $T\SI{<15}{\second}$ for a production or drilling semisubmersible used in hydrocarbon production) the heave response can generally be predicted satisfactorily through a linear potential flow computation. 
The contribution from vortex shedding is however particularly important at low frequencies around cancellation and resonance. Specifically, at cancellation, the drag force is the only form of excitation. 
At the resonance frequency, the drag constitutes the primary source of response attenuation, with the motion amplitude given by $s_{3a}=F_{3a}/\left(b_{33}\omega_R\right)$.

Wave basin tests are typically performed in order to estimate the viscous damping and to therefore derive empirically corrected RAOs, for instance, such as those reported by \cite{Ostergaard1987,MasSoler2018}. 
Notably, the model scale experiments for a conventional semisubmersible by \cite{MasSoler2018} demonstrate the dependency of the cancellation and resonant response on the wave amplitude. However, the majority of the studies reported in the literature are typically contextualised to establishing the response to design (extreme) seastate waves.

It is important to consider that experiments based on broadband (or white noise) wave energy, conventionally utilised for system identification, or design waves conditions with large amplitude, would likely not lead to a suitable characterisation of the expected heave response to swells in operational conditions. Testing in design waves, in particular, is likely to result in a significantly over-damped response when applied to operational seastates and therefore non-conservative motion predictions. 
It is generally not feasible in industry to undertake a large set of experiments with varying underlying excitation of wave spectra representative of low to moderate height swells in order to fully characterise the response. 


\subsection{Surface wave forecasts}
With respect to the wave energy providing the underlying forcing the vessel response, third generation spectral wave models are the state of the art for capturing wave generation and tracking swell propagation across ocean scale and local scale distances. Models such as WAM \citep{Wamdi1988}, WaveWatchIII (WW3, \cite{Tolman2014}) and SWAN \citep{Booij1999} form the basis of many operational forecasting products used globally for the purpose of forecasting wave conditions in the offshore and nearshore environment. The directional wave spectra at specified time horizons are outputs from the models. 

Coverage on the fundamentals and development of Numerical Weather Prediction which forms the basis of these models is provided by \cite{Brotzge2023}, for instance. In brief, the wave models solve the spectral action density balance equation for wavenumber-direction spectra based on a parameterisation of the source and dispersion terms. The atmospheric wind forcing is provided by numerical weather prediction models. 

Computational enhancements have enabled the main forecasting institutions to run these spectral wave models operationally at higher spatial resolutions over the years. This has captured more of the dynamics present in regional and local-scale systems that drive wave growth and decay. However, the spectral frequency resolution has continued to present a key constraint. Specifically, the logarithmic scaling of the frequency domain that these models adopt means that the region of sensitivity in the response of most semisubmersible vessels is sparsely resolved. 

Furthermore, while these models have proven to be accurate in observing peak conditions, for operational planning these are often of secondary importance to the models ability to capture the onset and decay of significant events and numerical wave models often struggle to correctly capture the timing of swell events. This is due, in part, to a lack of comprehensive observations to data assimilate across the extremely long fetches over which swells propagate and also the lack of resolution across these lower frequencies that induce numerical error. 

\subsection{Vessel response forecasts}

Herein, the vessel motion forecasts are considered in the context of a conventional twin-pontoon, four column, dynamically positioned semisubmersible drilling rig. The vessel operated in the North West Shelf of Australia in a water depth of around \SI{350}{\meter}. The location is characterised by a generally bimodal sea state, comprising locally generated short-period wind sea and long period swell during the Austral winter originating from storms in the Southern Ocean. While the location is also affected by tropical cyclones during the summer months, they were not present during the time period analysed in this study.


The input constituents to the physics-based forecast model of vessel response comprise the hydrodynamic RAOs and the spectral wave energy forecasts. 
A single set of hydrodynamic RAOs were available, corresponding to the nominal operational draft of the semisubmersible of approximately \SI{20}{\meter} and displacement of around \SI{46000}{mt}.

Spectral wave energy forecast data for the location were sourced from AUSWAVEG3, a deterministic numerical wave model run by the Australian Bureau of Meteorology. The model and its operational set-up is documented by \citep{Zeiger2021}. In brief, model is based on version 6.07 of the WW3  spectral wave model and is forced with the ACCESS atmospheric model. That is a global data assimilating atmospheric model with a horizontal resolution equivalent to \SI{12}{\kilo\meter}. AUSWAVEG3 has a base resolution of \SI{0.125}{^\circ} (\SI{12.5}{\kilo\meter}), increasing to \SI{0.0625}{^\circ} (\SI{6.25}{\kilo\meter}) around islands and in regions of less than \SI{350}{\meter} depth. The model was run multiple times a day, providing forecast output at hourly intervals up up 10 days from the time of issue. 

The output provided by the AUSWAVE G3 model is the full directional energy spectral density, discretised into 30 azimuthal directional bins of equal width and 28 frequency bins logarithmically spaced from 0.0412 to \SI{0.5399}{\hertz}. As mentioned, while this logarithmic scaling results in the shorter wave periods being accurately resolved, the resolutions of lower frequencies where the vessel is most responsive (nominally between 17 and 20 seconds) are sparsely resolved.



With these inputs, statistics of the vessel response can be readily computed using a physics model. Following, for instance, \cite{Faltinsen1990}, the spectral moments of the displacement is given by 
\begin{equation}
\label{eqn:mi}
    m_i = \int_0^\infty\int^{2\pi}_0 \omega^i|RAO|^2S\left(\omega,\theta\right) \, \mathrm{d}\omega \, \mathrm{d}\theta.
\end{equation}
The significant response amplitude can be estimated directly from the zeroth spectral moment, where $s_{a,1/3}=2\sqrt{m_0}$. We emphasise that it is this statistic which we proceed to subsequently apply a correction to using the Bayesian linear regression. 

In practice, due to the discrete nature of the forecast and RAO data, equation~\ref{eqn:mi} is approximated as $m_i=\sum \omega^i S\left(\omega,\theta\right) \Delta \omega \Delta \theta$, where $\omega$ and $\theta$ are typically the centre frequencies and directions. Further, the frequency resolutions of the wave energy and RAOs are generally not equivalent, necessitating an interpolation when used to establish a deterministic response forecast. Given the relatively large frequency bins of the wave energy at low frequencies, in order to ensure the strong resonant characteristics are preserved, an appropriate strategy would appear to be based on an interpolation of the wave energy to the RAO data.

\begin{figure*}
    \centering
    \includegraphics[scale=0.6]{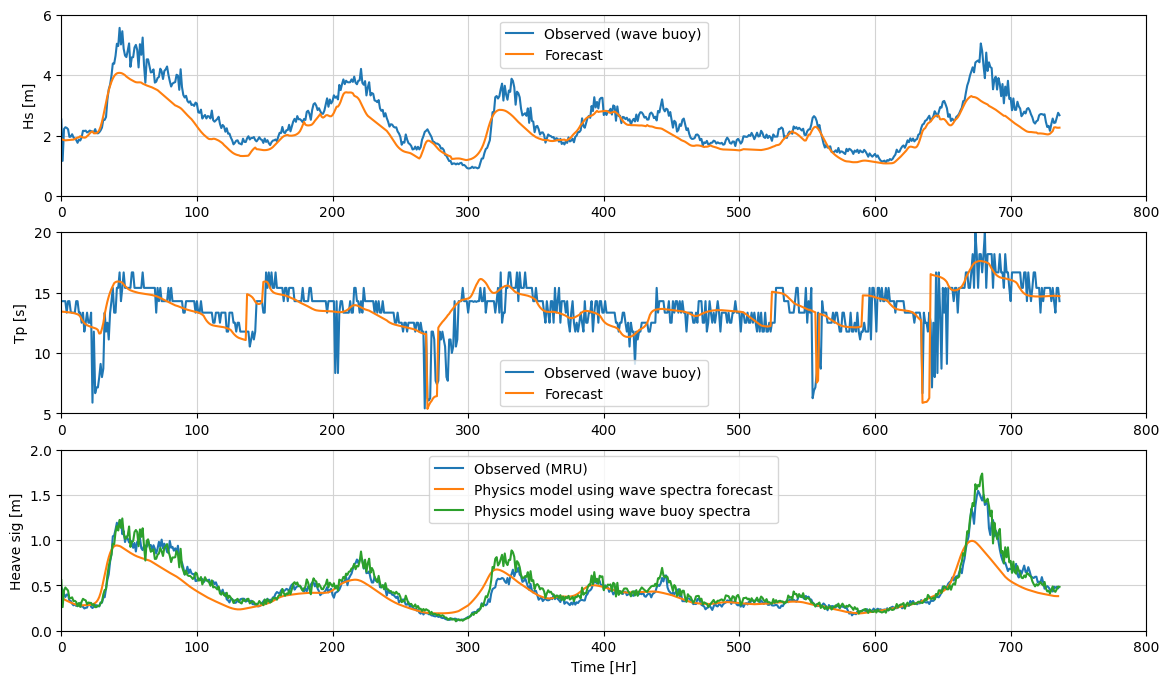}
    \caption{From top: time histories of the significant wave height (Hs), peak spectral wave period (Tp) and significant heave displacement ($2\sqrt{m_0}$). Comparisons are shown between the observed and predicted values (from the numerical forecast or, for the heave motion only, the wave buoy).
    Results are at an \SI{1}{\hour} time step over a 1 month period.}
    \label{fig:data_physics}
\end{figure*}


\subsection{Efficacy of a pure physics model forecast}
To motivate the need for a hybrid model approach, we now demonstrate the performance of a pure-physics model when using wave spectra sourced from a raw numerical forecast; in this case AUSWAVEG3. Figure~\ref{fig:data_physics} shows predictions of the zero-hour (i.e. nowcast) significant heave displacement for the semisubmerible rig, as compared to the measured values. The data comprise a 1 month period in which around 10 swell events were observed.
The results show that while the predictions are in general agreement with the measurements, the larger heave responses appear to be systematically under-predicted.

To elucidate the source of these discrepancies, predictions from the physics model were alternatively generated using wave spectra measured by a nearby wave buoy. These results are also plotted in Figure~\ref{fig:data_physics}, and which align much closer with the measured motions. This demonstrates that the physics and RAOs underlying equation~\ref{eqn:mi} are not in obvious error, implying that the wave forecast is a dominant source of the disparities. 

To support this hypothesis, corresponding forecast and measured values of the significant wave height and peak spectral period are also plotted. These show that the events where errors in the heave were relatively large, the $H_s$ was generally underpredicted by the forecast. The largest error aligned with a $T_p$ close to the heave resonance of rig. The fact that errors persist even for the nowcast is attributed to a lack of assimilation of wave buoy data in forecasts for this region, notably due to the large fetches involved. 

\section{Bayesian forecasting, probabilistic scoring metrics and the hybrid model}\label{sec:hybrid}
This section begins with an introduction to Bayesian statistics, with an emphasis on probabilistic forecasting.  The fundamentals of Bayesian statistics relevant to this work are presented and for deeper expositions on these topics the interested reader is directed to the comprehensive accounts in   \cite{bernardo2009bayesian} and \cite{gelman2013bayesian}. The important topic of validation and comparison metrics, known as proper scoring rules when considering probabilistic forecasts  \citep{gneiting2007probabilistic} is then discussed.  Finally, we introduce the baseline hybrid model that involves embedding a time evolving forecast of vessel heave response within a Bayesian statistical framework.  

\subsection{Bayesian statistics and probabilistic forecasts}
Assume we have a measurement, $y_1$, that is a realisation of a probabilistic data generating mechanism $p(y|\theta)$, where $\theta$ contains the collection of unknown model parameters. Bayesian statistics specifies a {\em prior} probability distribution, $p(\theta)$, that encodes our knowledge of $\theta$. A prior distribution may be more or less informative.  Physical knowledge of the system and/or expert opinion may serve to inform $p(\theta)$, thereby reducing our uncertainty (see \cite{Astfalck2018} for examples in the context of ocean engineering).  If prior knowledge of $\theta$ is scarce,  noninformative or ``objective'' priors  may be chosen to reflect a greater uncertainty regarding $\theta$ \citep{BERNARDO2011263}.   

Initially, the goal is to update $p(\theta)$ with information provided  $p(y_1\mid \theta)$ to $\theta$'s {\em posterior} probability distribution, denoted by $p(\theta\mid y_1)$.  Bayes Theorem permits a coherent "uncertainty accounting" action to do so by inverting the conditional direction in the numerator of the standard conditional probability rule: 
\begin{equation} \label{eqn:bayes}
     p(\theta \mid y_1)  =  \frac{p(y_1 \cap \theta)}{p(y_1)} = \frac{p(y_1 \mid \theta) p(\theta)}{p(y_1)}
\end{equation}
where $p(y)=\int p(y_1,\theta)d\theta$ is the normalising constant of $p(\theta|y_1)$ and does not depend on $\theta$.  Equation \eqref{eqn:bayes} thus offers a fully probabilistic account of our uncertainty associated with model parameters, given our prior and data model.  Extending our interest to forecasting a future unknown value $y^\ast_2$ (where the asterisk signifies $y_2$ is yet to be observed), from Equation \eqref{eqn:bayes} the {\em posterior predictive} distribution of $y^\ast_2$ is available as 
\begin{eqnarray}\label{eqn:predictive}
    p(y^\ast_2 \mid y_1) & = & \int p(y^\ast_2,\theta \mid y_1) \; \mathrm{d}\theta \nonumber \\   & = & \int p(y^\ast_2 \mid \theta) p(\theta \mid y_1) \; \mathrm{d}\theta 
\end{eqnarray}
from which we may obtain  probabilistic  intervals to an arbitrary degree of precision.  

Two important points are worth noting in Equations \eqref{eqn:bayes} and \eqref{eqn:predictive}.  First,  the posterior predictive of the unobserved $y_2^\ast$ is conditioned on the observed $y_1$ only, with $\theta$ integrated out such that $p(y^\ast_2|y_1)$ is the posterior expectation of $p(y^\ast_2 \mid \theta)$ and accounts for all uncertainties associated with the given statistical model. Second, were $y_2$ to become observed in future, it can then be used in the same manner to obtain the posterior predictive distribution of $y_3^\ast$, $p(y_3^\ast|y_2,y_1)$.  In theory, this process can continue indefinitely: write $y_{1:t}=(y_1,y_2,\ldots,y_t)$ and denote by $h$ a forecast horizon of interest, then we may continue to update $p(y^\ast_{t+h}|y_{1:t})$ for $t=1,2,\ldots$.    

As a final note; typically,  the normalising constant, for example $p(y_1)$ in Equation \eqref{eqn:bayes}, is intractable and hence  $p(\theta|y_1)$ must be numerically estimated.  We implement a Hamiltonian Monte Carlo algorithm in PyMC3 to draw samples from $p(\theta|y_1)$ \citep{salvatier2016probabilistic}.

\subsection{Proper scoring rules and validation metrics}\label{sec:scoring}
Scoring rules are numerical summaries that involve both the predictive (in our case, posterior) distribution and the value that actually materialises.  A proper scoring rule is one that is maximised if the the predictive distribution used is the true distribution. Following from the previous subsection, if $y_2$ were to become available, it not only updates our information to forecast $y^\ast_3$, but can serve to compare $p(y_2^\ast\mid y_1)$ with the value $y_2$ takes.  As we proceed into the future we can continue to update and compare forecasts to identify the optimal predictive procedure for a given scoring rule. The below is a summary of \cite{Astfalck2023}'s discussion of scoring rules in the context of offshore engineering.

There are many scoring rules, of which this article includes two that can serve for both probabilistic and non-probabilistic forecasts in order to compare the hybrid model with the physics-based deterministic forecast such as the one described in section \ref{sec:phys_approach}. With non-probabilistic forecasts, accuracy of the forecast is often sufficient and is assessed using root mean square error (RMSE).  RMSE is a measure of accuracy that is intuitive, commonplace in engineering and, as it turns out, a proper scoring rule. Not so commonly addressed in the engineering literature is the precision of a forecast is also important, whereby precision  we mean widths of given probability intervals that represent the uncertainty associated with the forecast.  If the intervals are too wide the model becomes operationally useless or if the intervals are spuriously narrow then operators may become overly confident in their decision making. One way to assess the correctness of a forecast's precision is its continuous rank probability score (CRPS): a proper scoring rule designed to assess both accuracy and precision, concurrently, and can be used for both probabilistic and deterministic forecasts \citep{leung2021forecast}.  

To be more precise, define $F_{t,h}$ as the predictive model at time $t$ predicting the process at horizon $h$. The RMSE is calculated as an aggregate score, over times $t$, as
\begin{equation}
    \mathrm{RMSE}_{h} = \frac{1}{T}\sum_{t=1}^T (\mathrm{E}[F_{t,h}] - y_t)^2
\end{equation}
where $\mathrm{E}[F_{t,h}]$ is the expectation of $F_{t,h}$. When the expectation is not analytically available it can be calculated as $\mathrm{E}[F_{t,h}] = n^{-1} \sum_i y^\ast_{t,i}$, where the $y^\ast_{t,i} \sim F_{t,h}$ are $n$ Monte Carlo draws from $F_{t,h}$. 

\begin{figure}[htp!]\label{fig:heaviside}
\centering
\includegraphics[scale=0.8]{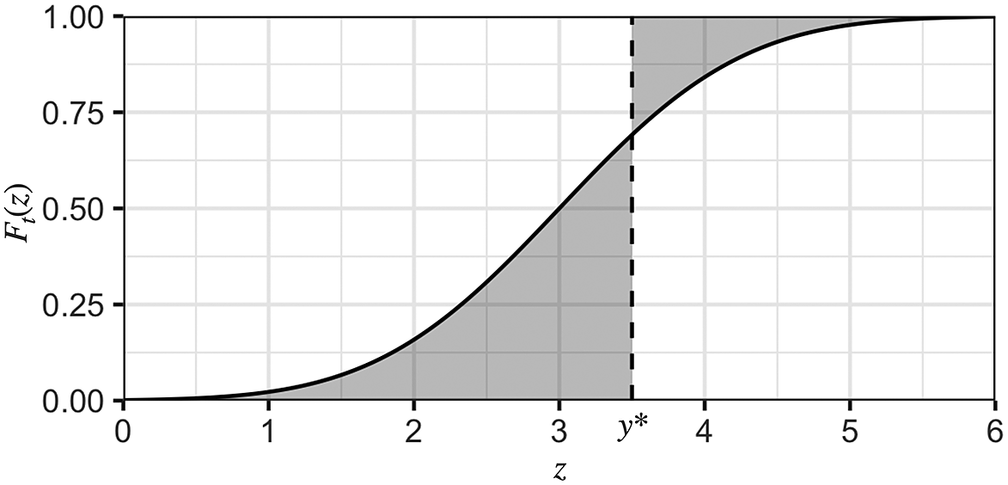}
\caption{Graphical interpretation of the CRPS. The distance in the integral in Equation (9) is represented by the shaded area. Forecast  $F_t(z)$
  is a cummulative distribution function -valued quantity.}
\end{figure}

The CRPS is calculated from the integral
\begin{equation}
\text{CRPS}_{t, h} = \int (F_{t,h}(z) - \mathbb{1}\{y_{t} \geq z\})^2 \; \mathrm{d}z.
\end{equation}
where $\mathbb{1}$ is an indicator function that returns a value of $1$ if the conditions in the bracket are met, and $0$ otherwise. The CRPS computes the distance between the cumulative density function $F_{t,h}(z)$ and the observation modelled as a Heaviside step function; see Figure~\ref{fig:heaviside} for illustration.

For certain analytical choices of $F_{t,h}$, the CRPS is analytically calculated \citep[see][]{gneiting2007probabilistic}. Alternatively, when Monte Carlo samples from $F_{t,h}$ are available, the CRPS is calculated as
\begin{equation}
    \text{CRPS}_{t,h} = \frac{1}{m} \sum_{i=1}^m |y^\ast_{t,i} - y_t| - \frac{1}{2m^2} \sum_{i=1}^m \sum_{j=1}^m |y^\ast_{t,i} - y^\ast_{t,i}|,
\end{equation}
where as before, $y^\ast_{t,i}, y^\ast_{t,j} \sim F_{t,h}$.
 
Both RMSE and CRPS will be used to compare the physics based approach with our hybrid model across multiple forecast horizons in section \ref{sec: diag_res_comp}.

\subsection{The hybrid model}\label{sec:linear_adj}

Research and frameworks based on a hybrid, (physics and data-driven) techniques have attracted increasing interest in offshore engineering in the past decade, allowing for probabilistic predictions informed from observed measurements that can be constrained by the underlying physics. Retaining access to the physics aids interpretation of the results and may lead to fundamental new insights into the underling mechanistic phenomena. Particular to Bayesian methods in the offshore engineering literature, \cite{astfalck2019emulation,Astfalck2019} present emulators for vessel motion models that retain the physics encoded in complex computer models, while providing uncertainty quantification and computational efficiency,  \citep{Loake2022} discuss updating metocean forecasts (in particular, wind) with linear regression in the presence of residual dependencies, \cite{sarkar2018prediction,sarkar2019spatiotemporal} consider Gaussian process models of surface currents and \cite{manderson2019} present a Bayesian hierarchical model for non-linear internal waves.

While the physics-based approach outlined in Section~\ref{sec:phys_approach} provides deterministic-based forecasts of vessel response, each of those predictions are point estimates that would benefit from a statistical quantification of uncertainty. Uncertainties arise from multiple sources including imprecision in the wave forecasts and in the RAO, which could exist across any of the amplitude, direction or temporal axes.  Here, the physics-based forecast is extended to a hybrid-forecast to serve two purposes: first, in an attempt to quantify the propagation of uncertainties arising from lower level measurements, and second, to `correct' the purely physics-based predictions if systematic discrepancies between predictions and observations exist. 

Figure \ref{fig:data_physics}, section \ref{sec:phys_approach},  suggests an accurate description of significant response amplitude is obtained using the zeroth spectral moment, $2\sqrt{m_0}$, obtained from the discretised version of equation \eqref{eqn:mi}.  We embed this physical model into the data generating mechanism as follows. Denote by $x_{t+h\mid t}=2\sqrt{m_{0,t,h}}$ a forecast for a given horizon, $h$, issued at time $t$,  accompanied by a heave measurement $y_{t+h}$ and write
\begin{equation}\label{eqn:linear_adjust}
   y_{t+h} = \beta_{0,h} + \beta_{1,h} x_{t+h\mid t} + \epsilon_{t+h},
\end{equation}
where $\epsilon_{t+h}$ is the corresponding error term, and $\beta_{0,h}$ and $\beta_{1,h}$ are the regression coefficients, which we define as bias and scaling adjustments of the forecast $x_{t+h\mid t}$.  In addition, Figure \ref{fig:data_physics} and domain expertise suggest that typically $2\sqrt{m_0}$ is not biased, or at least not obviously so.  Consequently, we set prior for the bias to be Gaussian with mean zero and variance 3: $p(\beta_{0,h})=\mathcal{N}(0,3)$.  It is also highly unlikely  for the scale adjustment to be be negative as this would imply output from the physical model that is in direct contrast to the measurements and we put a Gaussian prior on the scale with mean 1 and variance 3 but truncated to be above zero: $p(\beta_{1,h})=\mathcal{TN}(1,3)$.
 
The structure of error term in equation \eqref{eqn:linear_adjust}, $\epsilon_{t+h}$, is important to consider because if assumptions of normality, independence and constant variance are obviously violated then forecasts derived from the posterior predictive distribution will be misleading.  We turn to this issue in section \ref{sec:diagnostics} which impels the hybrid model of equation \eqref{eqn:linear_adjust} to be extended to include a more complex error structure.

\section{Data}
\label{sec:data}
\subsection{Measurements of vessel heave}
Key to enabling the hybrid-forecast are measurements of the heave displacement of the semisubmersible Mobile Offshore Drilling Unit (MODU) at the centre of gravity (equivalent to the RAO origin). These data were acquired using motion reference units (MRUs) which sampled continuously at a rate of \SI{1}{\hertz}. The measurements from multiple MRUs that were positioned at different locations on the vessel which were found to be in close agreement and provided assurance of the high quality of the data. 

Marine reports detailing the hydrostatic properties were also available and used to monitor the operating condition vessel (in particular the centre of gravity). The quality-assurance process involved identifying and excluding instances where the MODU was transiting, changing heading or mass (draft). 

A high-pass digital filter ($5^{th}$-order Butterworth) was subsequently applied to the MRU data using a cut off frequency of \SI{0.4}{\hertz}. This was on the basis that slow drift effects were not able to be modelled using linear wave theory and that the lowest frequency bin of the wave spectra forecasts was \SI{0.0412}{\hertz}. 
The zeroth moment statistics of the heave displacement were computed from these processed time histories, based on a \SI{3}{hour} averaging period and \SI{1}{\hour} sampling interval. 

\subsection{Compilation of predictions for forecast horizons and training-test data split}

Data sets comprising the raw model motion forecast corresponding to the date and time of the measurements were synthesized for different forecast horizons.
For the AUSWAVE G3 model considered here, the time interval between successive forecast issues was at least \SI{6}{\hour}. Therefore, to create a data set of continuous motions forecasts at \SI{1}{\hour} intervals, the predictions at all hours between the successive issues were concatenated. For instance, the data set which we refer to as $h=0$, was prepared by taking the predictions of the 00, 06, 12, 18Z forecasts spanning between the time of issue ($t$) and $t+\SI{5}{\hour}$ and concatenating the results into a continuous time series. Since the 06 and 18Z forecasts did not extend past \SI{72}{\hour}, the motion forecasts for the longer time horizons were based on only the 00Z and 12Z issues. That is, the $h=72$ data set comprised a concatenation of the forecasts between $t+\SI{72}{\hour}$ and $t+\SI{83}{\hour}$. Synethesising time histories in the manner described above inherently considers the most recent forecast information that would have been available to the end user.

The training and test data sets was formulated based on a 80:20 split. Since it is important to preserve the time synchronicity, the ensembles were not selected at random. The training data were used to assess the in-sample posterior predictives and the test data were used for the out-of-sample posterior predictives.

\section{Diagnostics and results} \label{sec: diag_res_comp}
In section \ref{sec: diag_res_comp},  we first present some diagnostics that  motivate an additional level of complexity required for the error structure, $\epsilon_{t+h}$,  in section \ref{sec:linear_adj}, equation \eqref{eqn:linear_adjust} and compare the two models' posterior distributions of the linear adjustments and posterior predictive variances across all forecast horizons, $h=0,6,12,24,48,72,96$.  Second, using the more comprehensive error structure, we report comparisons between the purely physics based model and our hybrid model, in terms of the probabilistic scoring rules described in section \ref{sec:scoring} and present forecast outputs graphically, with probability bounds that represent our uncertainty associated with the forecast.  

\subsection{Diagnostics}\label{sec:diagnostics}
The assumption of the linear adjustment model of equation \eqref{eqn:linear_adjust}
is that the error terms are independent and have an identical Gaussian distribution, with common variance. Our experience with metocean time series is that such assumptions are rarely satisfied, even when the mean processes is well modelled using a physics based approach. First, as the underlying data represent a stochastic process with structure through time, the error is likely to be correlated. For instance, if the physics-based model forecast under-predicts at time $t$ then it is likely the model may also under predict at time $t+1$. This is particularly salient for swell modelling where one of the largest known errors arises from the limitations of numerical wave models to track swell pulses and even with a linear adjustment this issue is unlikely to be resolved. Further, the error variance is not expected to be constant as the magnitude of the error terms is typically larger in higher energy seas, and smaller in lower energy seas, particularly due to aforementioned frequency characteristics of the heave RAO. 

\begin{figure*}
\centering
\begin{tabular}{ccc}
    \includegraphics[scale=0.35]{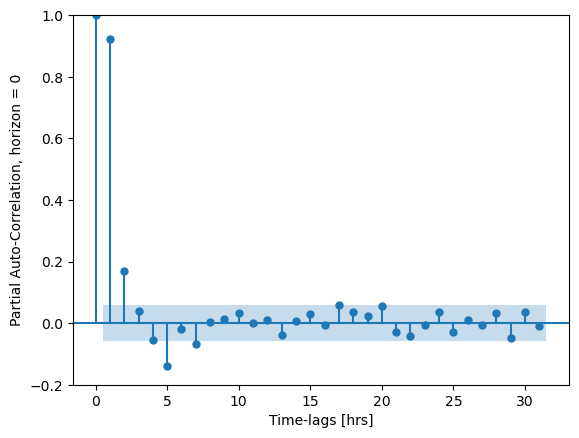} & 
    \includegraphics[scale=0.35]{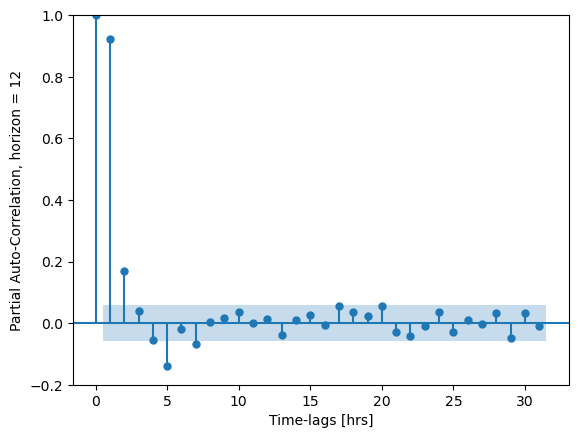} & \includegraphics[scale=0.35]{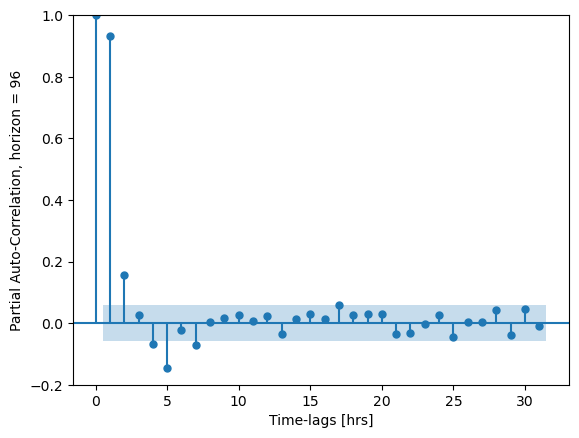}\\
    \includegraphics[scale=0.35]{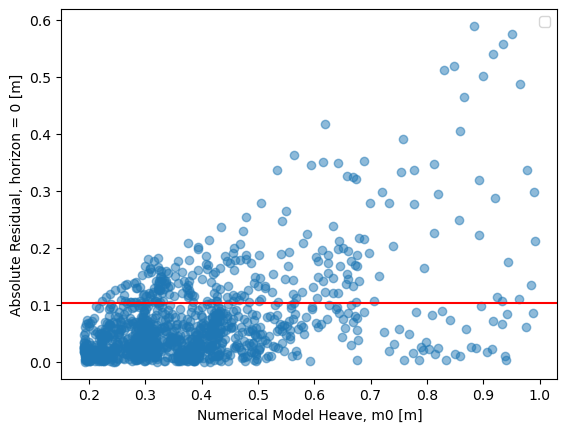}&
    \includegraphics[scale=0.35]{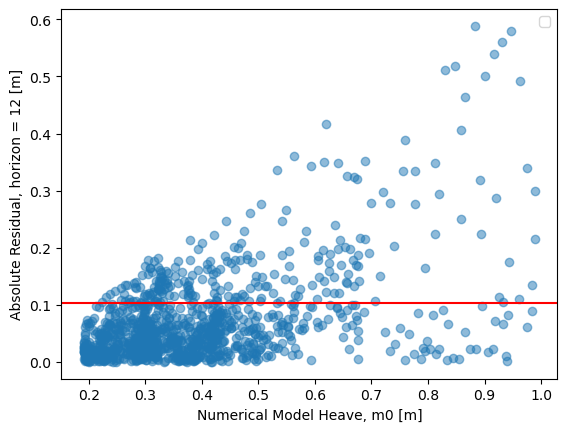}&
    \includegraphics[scale=0.35]{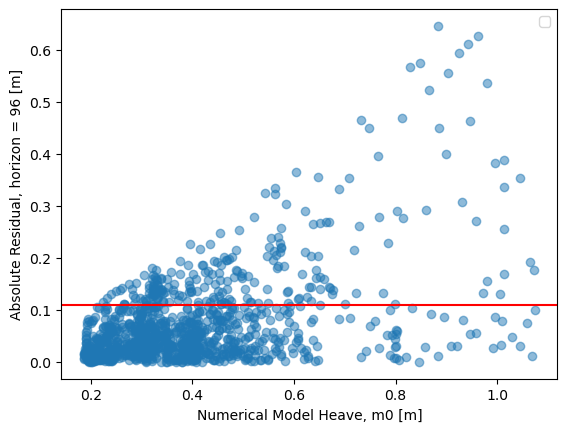}\\
\end{tabular}   
    \caption{Autocorrelation plots (upper panels) and absolute residuals versus numerical model heave (lower panels) for the linear adjustment model of equation $\eqref{eqn:linear_adjust}$. 
    From left to right: 0, 12 and 96 hour horizons.}
    \label{fig:h00_12_96_diagnostics_simplelinear}
\end{figure*}

We run the model of equation \eqref{eqn:linear_adjust} for forecast horizons  $h=0,6,12,24,48,72$ and $96$ hours. 
Figure~\ref{fig:h00_12_96_diagnostics_simplelinear} shows the error structure, in terms of partial autocorrelations (PACFs) of the residuals and (absolute) residuals versus the physical model forecast for $h=0, 12$ and $96$ (similar diagnostic plots for the other time horizons are contained in the appendix). For each time horizon, the structure is similar: the PACF plots report a residual structure that follows an autocorrelation process of order 2 (i.e., AR(2)) and the absolute residuals indicate that the variance of the residuals increases as the physical model forecasts contained in $x_{t+h|t}$ increase.  The red line in these latter plots are the {\em maximum a posteriori} (MAP) estimate of $\sigma$, the residual variance of equation \ref{eqn:linear_adjust} and clearly does not represent this increase variance of the residuals.  Such deviations from independence and homogeneity of the residuals compel us to extend the error structure of \eqref{eqn:linear_adjust}.  We proceed as follows: write 
\begin{equation}\label{eqn:linear_adjust_errors}
   y_{t+h} = \beta_{0,h} + \beta_{1,h} x_{t+h\mid t} + \phi_1\epsilon_{t-1} + \phi_2 \epsilon_{t-2} + x_{t+h\mid t} \eta_{t+h},
\end{equation}
where
\begin{eqnarray*}
\epsilon_{t-1}&=&y_{t-1}-\beta_0-\beta_1 x_{t-1\mid t-1-h}, \\
\epsilon_{t-2}&=&y_{t-2}-\beta_0-\beta_2 x_{t-2\mid t-2-h}, \quad \text{and}\\
\eta_{t+h} &\sim &N(0,\sigma^2_h).
\end{eqnarray*}

The inclusion, in equation \eqref{eqn:linear_adjust_errors}, of the lagged error terms, $\epsilon_{t-1}$ and $\epsilon_{t-2}$, and their coefficients, account for the AR(2) process and note that each error term is scaled by the physical model forecast, $x_{t+h|t}$, to account for the heteroskedasticity. 

\begin{figure*}
\centering
\begin{tabular}{ccc}
    \includegraphics[scale=0.35]{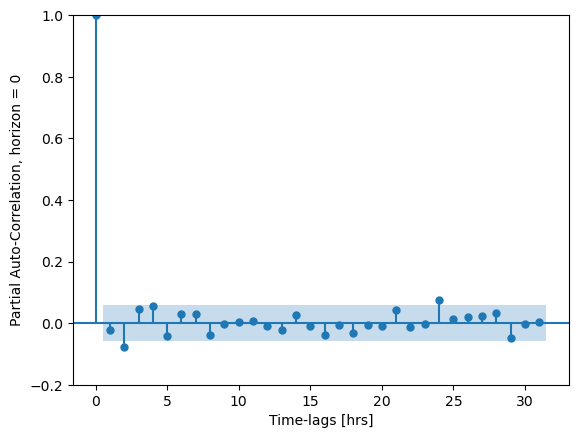} & 
    \includegraphics[scale=0.35]{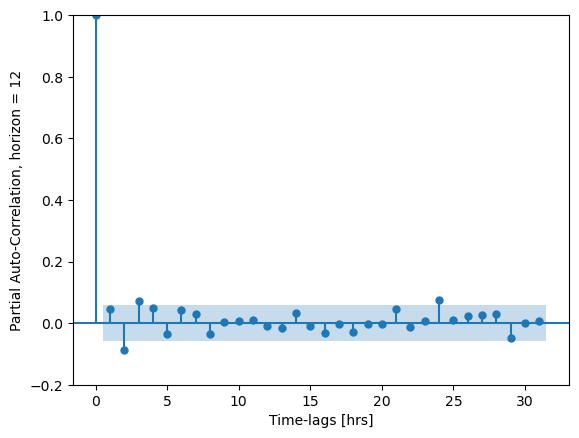} & 
    \includegraphics[scale=0.35]{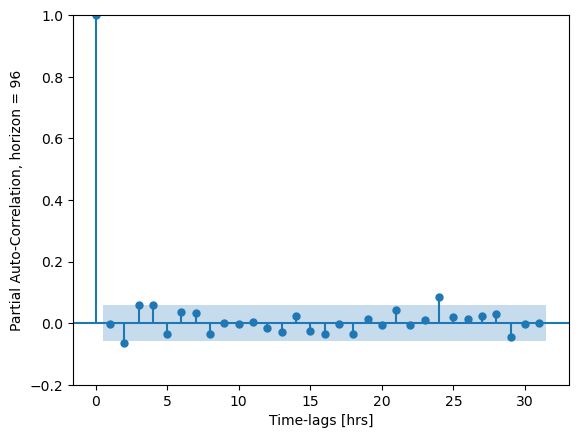}\\
    \includegraphics[scale=0.35]{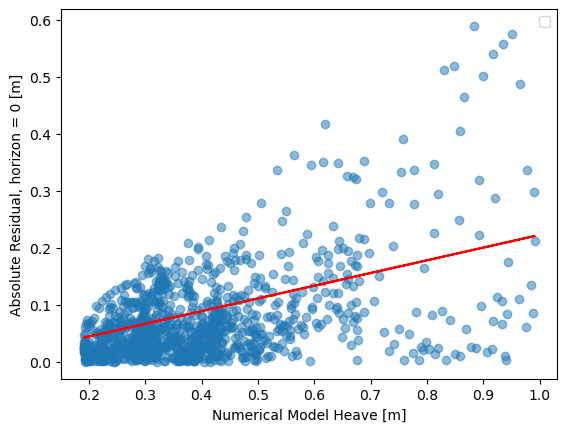} &
    \includegraphics[scale=0.35]{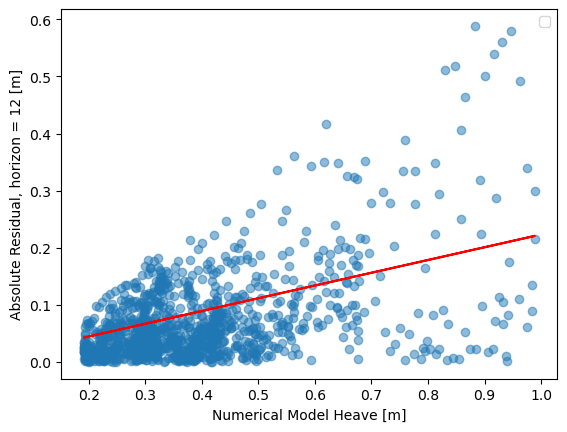} & 
    \includegraphics[scale=0.35]{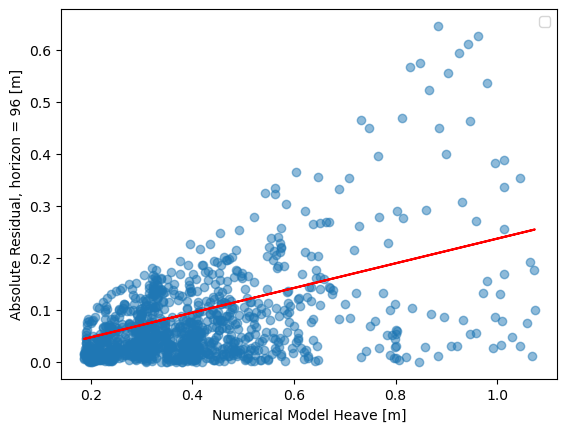}
\end{tabular}   
    \caption{Autocorrelation plots (upper panels) and absolute residuals versus numerical model heave (lower panels) for the linear adjustment model of equation $\eqref{eqn:linear_adjust_errors}$. 
    From left to right: 0, 12 and 96 hour horizons.}
    \label{fig:h00_12_96_diagnostics_linear_adjust}
\end{figure*}

The lower panels of Figure~\ref{fig:h00_12_96_diagnostics_linear_adjust} are similar to the upper panels, except they represented residual structure, $\eta_{t+h}$ in equation \eqref{eqn:linear_adjust_errors}. We note the residuals are now approximately uncorrelated and the MAP of $\sigma$ now reflects that the variance increases with $x_{t+h|t}$.  

All hybrid results presented in section \eqref{sec:results} are therefore based on equation \eqref{eqn:linear_adjust_errors}.

\subsection{Results}\label{sec:results}
\begin{figure*}
\centering
\begin{tabular}{c}
\includegraphics[scale=0.5]{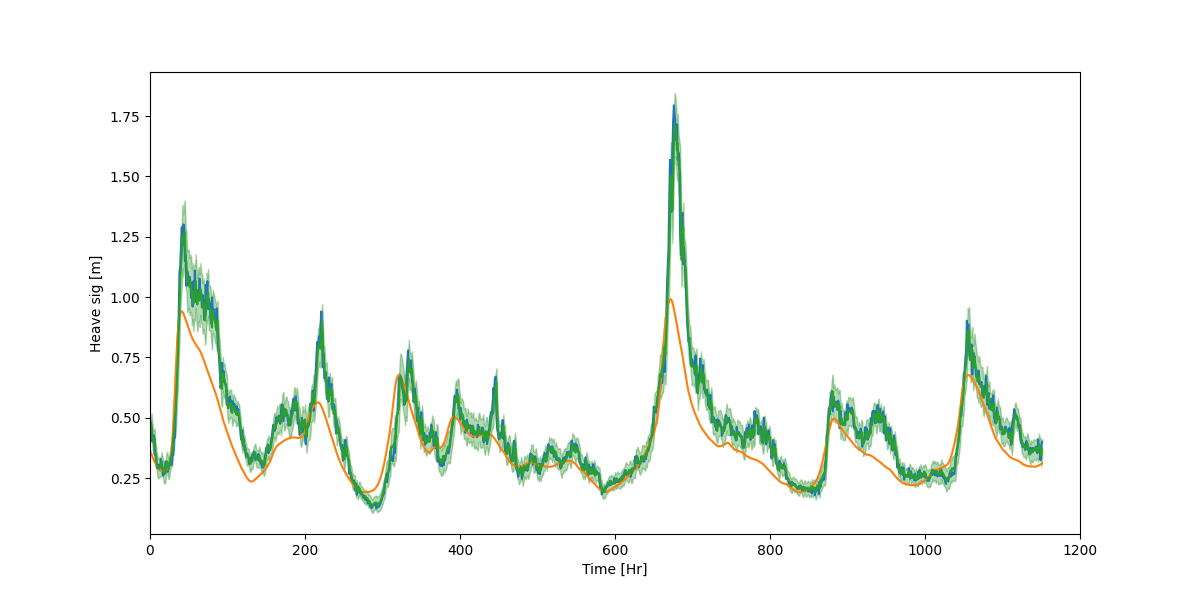}\\
\includegraphics[scale=0.5]{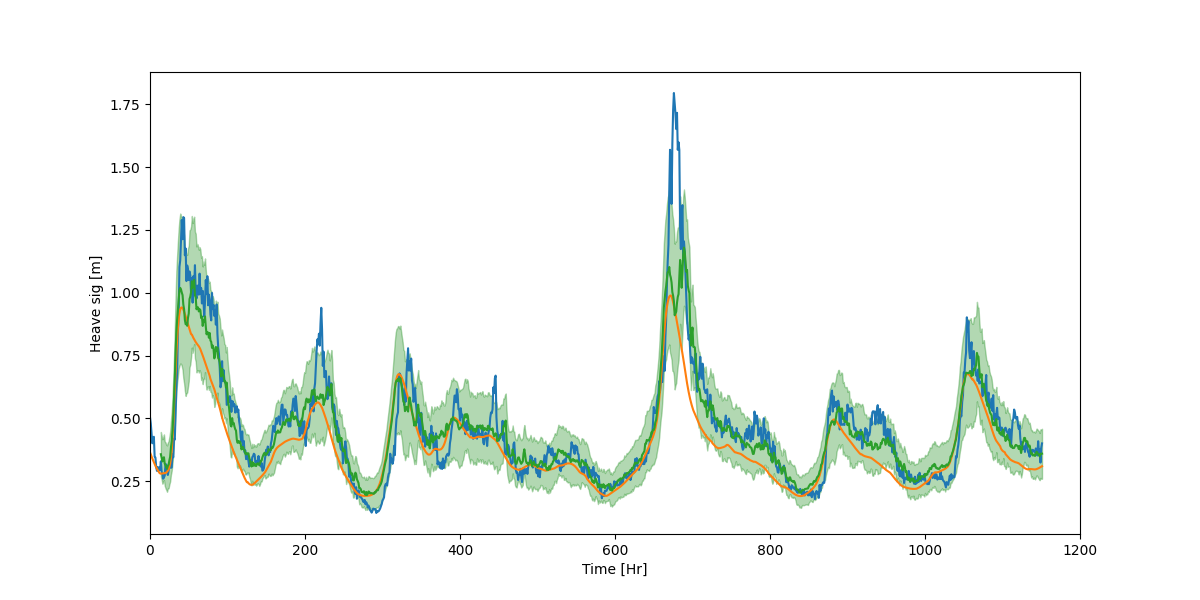} \\
\includegraphics[scale=0.5]{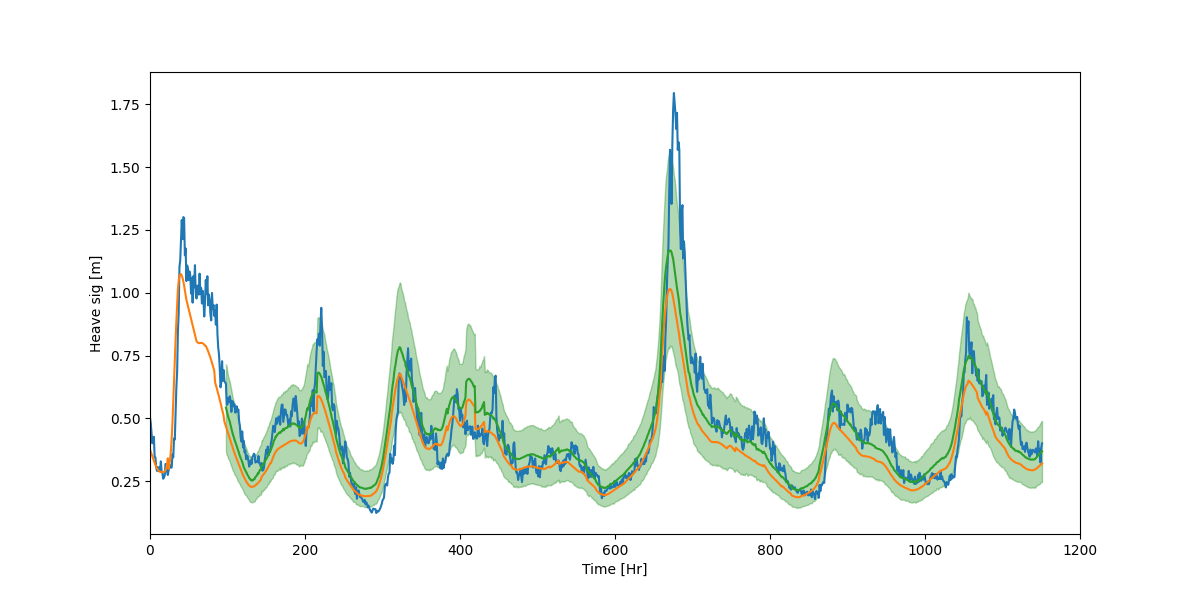} 
\end{tabular}
\caption{In-sample predictions for 0, 12 and 96 hour forecast. Shaded regions show the hybrid model's P5 and P95 credible intervals.}
\label{fig:Insample_prediction_timehists}
\end{figure*}

We now proceed to assess the performance of the hybrid model. First, we consider the in-sample posterior predictions for a period of data spanning approximately 2 months. Figure~\ref{fig:Insample_prediction_timehists} presents the time histories of the respective P50 predictions and P5 and P95 credible intervals for the time horizons of 0, 12 and \SI{96}{\hour}. The observed values and deterministic predictions of the raw physics-model are also shown for comparison. It can be noted that the raw model predictions do not vary appreciably between the different horizons. This characteristic is expected, given the lack of assimilation of wave measurements within the spectral wave forecasts for this region.

The hybrid model is demonstrated to provide superior predictions over the raw physics model, particularly at short horizons. Specifically, the hybrid model is shown to correct for a generally low-bias of the physics model, with the AR components being key enablers to this. While there are still underpredictions at the inceptions of the heave peaks (swell arrival), importantly, the AR terms are able to significantly improve the predictions thereafter. This implies that, at relatively short horizons, the measurements provide a significant adjustment to the raw physics model.

Given the strongly heteroskedastic nature of the data (and its assumed scaling with the heave magnitude), as is expected, for relatively low heave (low $H_s$ or relatively short $T_p$) the credible intervals are relatively small. In contrast, for the inception of the long period swell coinciding with large heave motion, the credible intervals are large. This is most apparent at long horizons for which the hybrid model becomes less responsive the measured data. However, in general, the credible bounds appear to provide a reliable measure of scatter in the predictions.

\begin{figure*}
\centering
\begin{tabular}{c}
\includegraphics[scale=0.5]{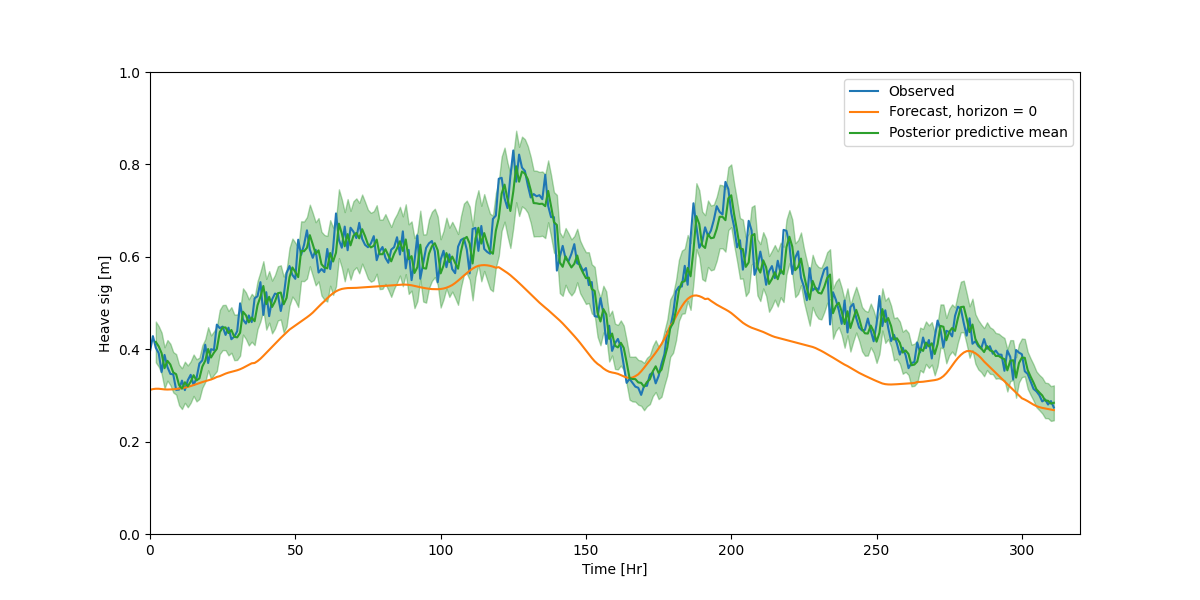}\\
\includegraphics[scale=0.5]{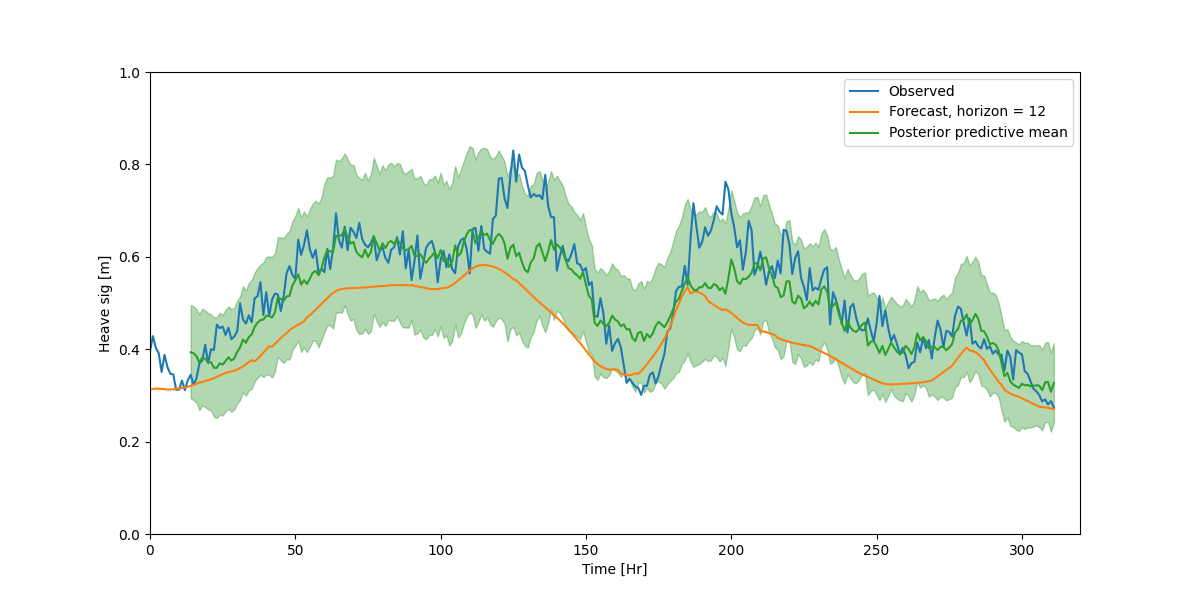} \\
\includegraphics[scale=0.5]{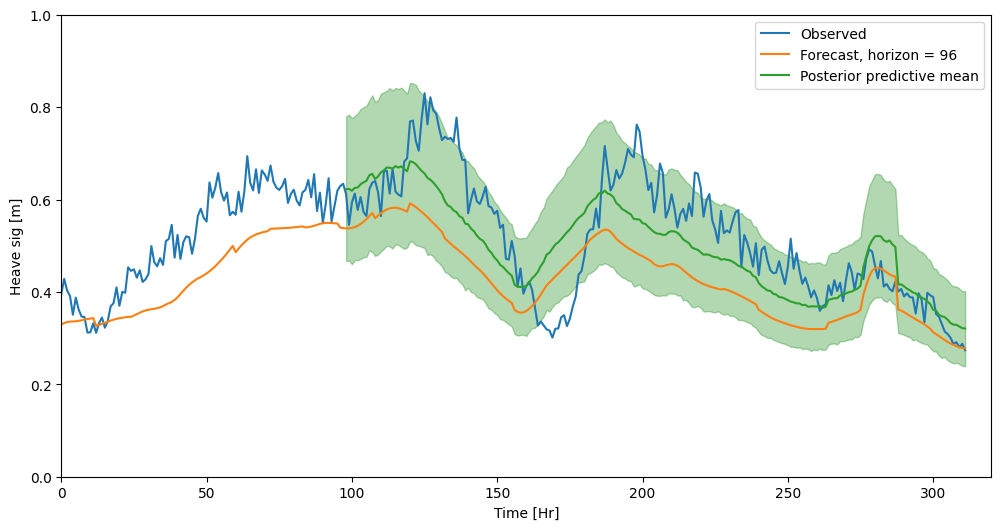} 
\end{tabular}
\caption{Out of sample predictions for 0, 12 and 96 hour forecast. Shaded regions show the hybrid model's P5 and P95 predictive credible intervals.}
\label{fig:Out_of_sample_prediction_timehists}
\end{figure*}

We now consider the out-of-sample tests, which are shown in and use the scoring rules to provide a quantitative evaluation of the models. Table~\ref{tab:outofsampletests} compares the RMSE and CPRS results between the hybrid model and raw model predictions for all forecast horizons. Given that the raw model forecasts were near invariant between horizons, it follows that both the RMSE and CRPS values were also near identical for all horizons. The hybrid model is shown to result in smaller values of RMSE and CRPS than the raw model at all horizons. It is emphasized that this demonstrates that the hybrid model not only yields predictions with smaller errors but, more importantly, enables more precise forecasts.  Figure \ref{fig:Out_of_sample_prediction_timehists} provides visual support of these scoring rules results: out-of-sample forecasts corresponding to $h=0, 12$ and $96$ show the hybrid model overwhelming outperforms the raw physics forecast.  Figure  \ref{fig:Out_of_sample_prediction_timehists} also demonstrates the hybrid model is able to provide out-of-sample P10 and P90 predictive credible intervals that the raw physical model cannot.

\begin{table*}
\caption{Out of sample test results (to 3dp)}\label{tbl1}
\begin{tabular}{l l l l l l l l l }
\toprule
&   & \SI{0}{\hour} & \SI{6}{\hour} & \SI{12}{\hour} & \SI{24}{\hour} & \SI{48}{\hour} & \SI{72}{\hour} & \SI{96}{\hour} \\ 
\midrule
RMSE& Bayesian model&  0.034 & 0.057 & 
0.069 & 0.079 & 0.074 & 0.078 & 0.089\\
    & Raw (pure-physics) model     & 0.119 & 0.12 & 0.119 & 0.118 & 0.118 & 0.119 & 0.117 \\
\addlinespace
CRPS& Bayesian model& 0.019 & 0.032 & 0.038 & 0.044 & 0.042 & 0.044 & 0.05\\ 
    & Raw (pure-physics) model     & 0.101 & 0.102 & 0.101& 0.100 & 0.099 & 0.102 & 0.099 \\ 
\bottomrule
\end{tabular}
\label{tab:outofsampletests}
\end{table*}

\section{Discussion}

\begin{figure*}
\includegraphics[scale=0.5]{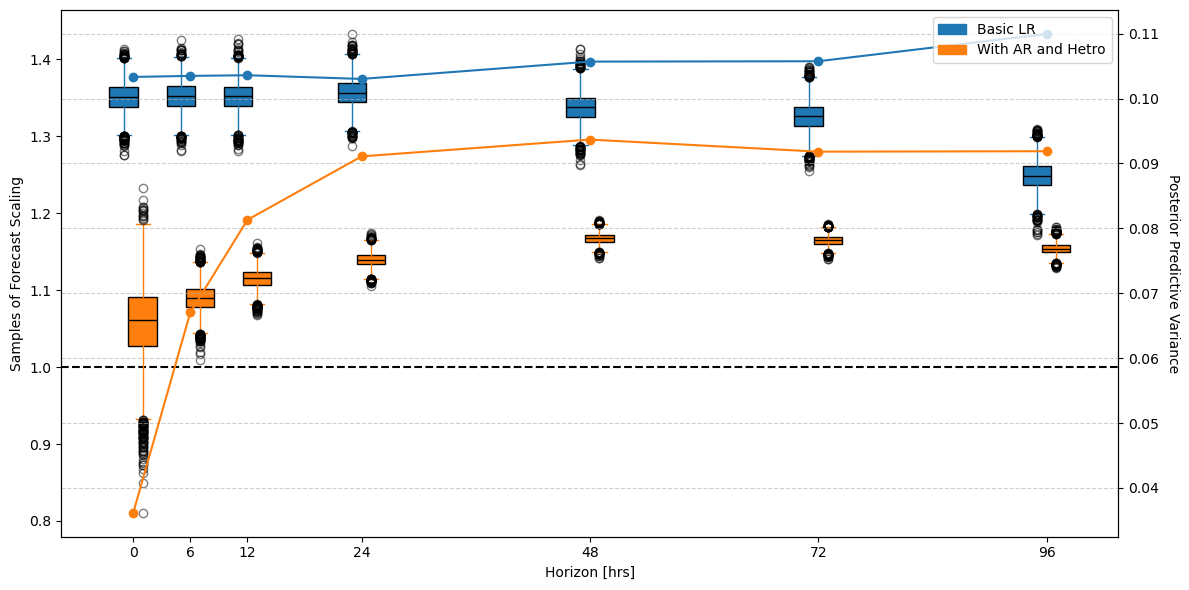}
\caption{Parameter summaries for $h=0,6,12,24,48,72$ and $96$ hour forecast horizons. Boxplots are estimated posterior distributions of the linear adjustment $\beta_{1,h}$ and solid dots are estimates of the predictive variance of $y_{t+h}$, $var(y_{t+h}|y_t)$.  Blue represents the basic linear adjustment model $\eqref{eqn:linear_adjust}$ and red represents the linear adjustment with heterogeneity and AR(2) process model $\eqref{eqn:linear_adjust_errors}$.}
\label{fig:eq9versuseq10}
\end{figure*}

The proposed framework systematically addresses total uncertainty in heave motion forecasting.
The method has been demonstrated to be effective for providing reliable quantification of uncertainty and correcting bias in the raw physics model predictions. As further evidence of this, it is interesting to quantify the adjustments made by the hybrid model of equation \eqref{eqn:linear_adjust_errors} to the data. Figure~\ref{fig:eq9versuseq10} shows the posterior distributions of $\beta_{1,h}$, which represent the bias correction of the raw physics model, across all horizons, together with estimates of the posterior predictive variance of $y_{t+h}$, $var(y_{t+h}|y_t)$. To understand the impact of the AR(2) and heteroskedastic terms in our hybrid model, corresponding results from the basic (physics-informed) linear model of equation \eqref{eqn:linear_adjust} are also provided for comparison.

First, in terms of $var(y_{t+h}|y_t)$ the hybrid model outperforms the basic linear model across all forecast horizons. This improvement is most prominent for shorter term forecasts where the AR(2) components are more relevant for the immediate future, but significant improvement is nevertheless maintained across further time horizons. This suggests the addition of heterogeneity and the autocorrelation structure to the basic linear model are beneficial for quantifying the uncertainty of the forecasts.  

Also, in Figure \ref{fig:eq9versuseq10}, both the basic linear model and the hybrid model demonstrate that the raw physics model requires a bias adjustment: $\beta_{1,h}=1$ indicates no bias correction is required for the raw physics model but, for all forecast horizons, the posterior distributions of the $\beta_{1,h}$s are significantly greater than 1. Notably, for all horizons, the $\beta_{1,h}$s of the hybrid model have smaller values compared to the basic linear model. This arises as the mis-specification of the noise process in the basic linear model over-estimates the regression parameters due to the occasional large swell events. This serves to demonstrate that the AR terms remain beneficial from a model specification point of view, even when their influence on prediction experience diminishing returns as the forecast horizon increases. Finally, for the hybrid model, at short horizons there is significant uncertainty in $\beta_{1,h}$, as it is not well-identified in the model: at very short horizons, the AR components serve to model the majority of the observed signal. At longer horizons the posterior variance in $\beta_{1,h}$ becomes much smaller, the bias correction becomes more important as the influence of the AR components wane and from $h=48$ onwards the posterior distributions of the bias correction stabilise.

Further rationale for the use of Bayesian linear regression over machine learning is that it retains a clear link to the underlying physics and remains robust with limited data. In comparison, more flexible machine learning methods require volumes of training data which would be unlikely available in practice. Our simple formulation provides a reliable, interpretable means of quantifying uncertainty and enables direct evaluation and progressive refinement of individual model inputs as better data become available.

\begin{figure*}
    \centering
    \includegraphics[width=0.8\linewidth]{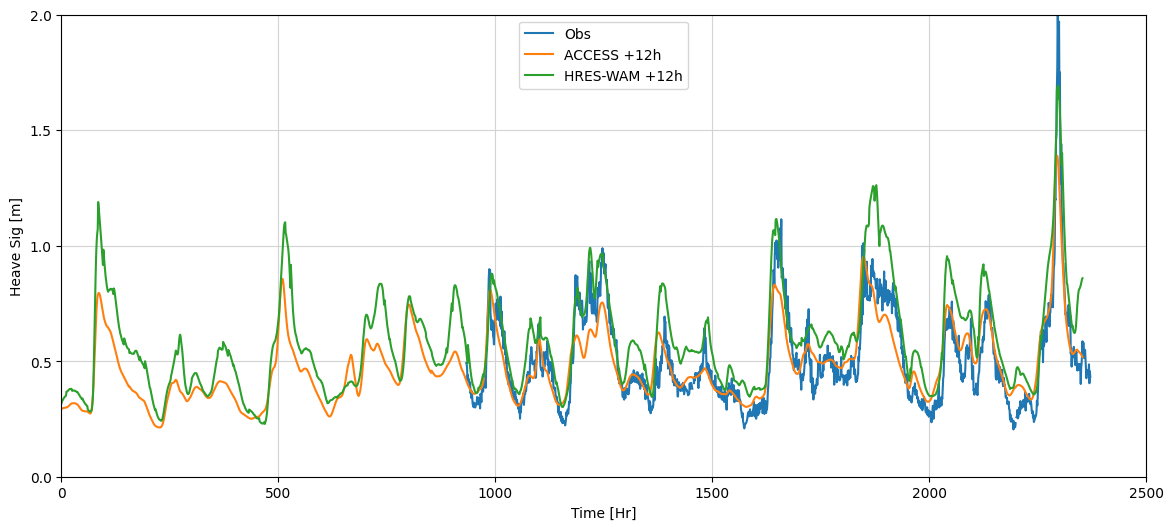}
    \caption{Physics-based forecasts using different spectral wave products compared to the measured heave response.}
    \label{fig:HeaveSig_Obs_BOM_EC_SCA}
\end{figure*}

To illustrate this, Figure \ref{fig:HeaveSig_Obs_BOM_EC_SCA} compares physics-based forecasts for a twin-pontoon semi-submersible generated using two spectral wave models (ACCESS and HRES-WAM~\citep{ECMWF2018}) and a common RAO dataset. The data correspond to a location similarly exposed to long-period swell and a nominal 12-hour forecast horizon. The pronounced differences between the two forecasts highlight the strong influence of the raw spectral input on overall motion error. Notably, these discrepancies manifest in both event peaks and troughs, the latter being particularly relevant for determining safe resumption windows for offshore operations following weather-related suspensions.

\begin{figure*}
    \centering
    \includegraphics[width=0.85\linewidth]{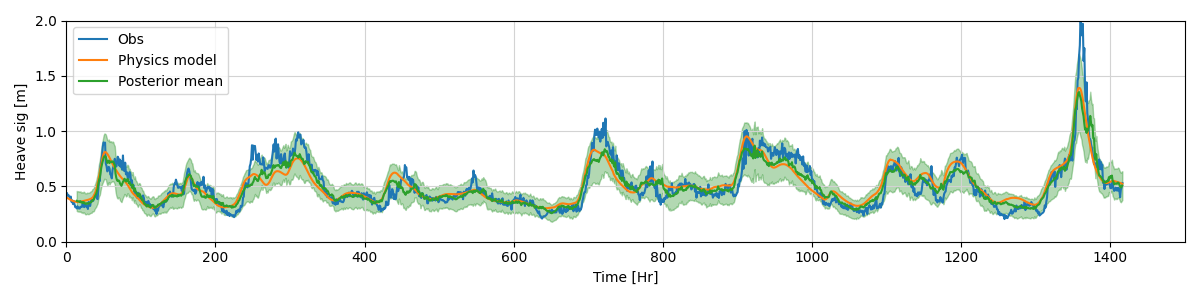}
        \includegraphics[width=0.85\linewidth]{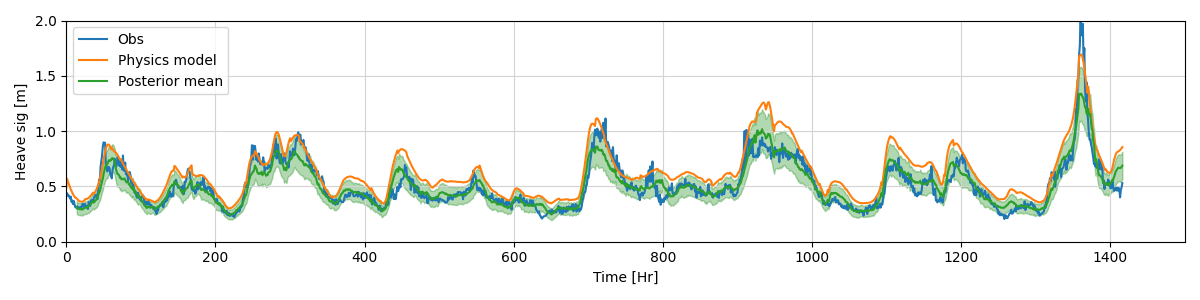}
    \caption{In-sample predictions for 12 hour forecast using the ACCESS (upper panel) and HRES-WAM (lower panel) spectral wave data. Shaded regions show the hybrid model's P5 and P95 credible intervals.}
    \label{fig:HeaveSig_Obs_BOM_LR12withAR_SCA}
\end{figure*}

Figure \ref{fig:HeaveSig_Obs_BOM_LR12withAR_SCA} presents the corresponding in-sample posteriors for the respective physics model predictions, obtained using the AR2 regression model with heterogeneous error treatment (the hybrid model of equation \ref{eqn:linear_adjust_errors}). The posterior means and associated credible intervals demonstrate good agreement with the measured data. The Bayesian adjustment is particularly effective for the HRES-WAM input, indicating a systematic high bias in the raw physics model. These results align with prior studies \citep{Milne2025} and reinforce the suitability and generalizable nature of the Bayesian framework for response-based forecasting.

Variability and systematic bias in spectral wave model outputs therefore remain key contributors to forecast uncertainty. Although ensemble spectral products are now common, their direct use in physics-based models may not yield reliable probabilistic forecasts, as the spectra themselves often contain systematic error. We reiterate, that these biases stem from limitations in swell propagation modelling, global wind forcing, and the assimilation of local observations, and are further influenced by model resolution. Continued advances in wave forecasting are expected to reduce these effects, and the Bayesian framework is well suited to incorporate such improvements.

The nonlinear (vortex) contribution to heave response makes reliance on a single RAO dataset also a potentially significant error source, particularly when long-period swell excites resonance or near-cancellation. It is reiterrated that over-damped RAOs can lead to non-conservative predictions and improved RAO fidelity is essential for borderline operational decisions where there are associated financial or logistical implications. 

Enchancement to the RAOs may be achieved through field measurement, model tank testing, or advanced numerical modelling of vortex damping. Field data overcome scale effects but require nearby wave buoys and long acquisition periods for statistically robust estimates, which may be impractical for short campaigns. Field-derived RAOs for semis have been reported by \cite{MasSoler2018} and \cite{Milne2024}, the latter showing the need for the need for strategic sampling near resonance to reduce estimation uncertainty. Wave-basin tests provide controlled conditions to quantify viscous damping and derive sea-state-specific RAOs, though long-period swell investigations require large basins and careful selection of model scale. Importantly, both field and basin data supply essential empirical evidence for developing and validating numerical models. While typically used to tune Morison damping coefficients, alternative formulations— based on isolated-edge theory—offer advantages in the low-KC regime by explicitly representing velocities around individual pontoon edges, addressing a key limitation of Morison-based methods.

\section{Conclusion}
This study has sought to address issues pertinent to the prediction of vessel motion with implications for safe and efficient offshore operations. Specifically, we have focused on the heave response of semis in long-period swells which is affected by multiple inherent challenges, namely predicting wave propagation over very large fetches and a vortex damped hydrodynamic resonance. A framework has been demonstrated to enable superior forecasts compared to conventional approaches. With access to measured heave motions, it treats for bias and provides probabilistic predictions to enable quantification of uncertainty. The inclusion of a heterogeneous and autoretrogressive treatment has been particularly justified for short forecast horizons. An advantage of adopting a Bayesian approach is that it maintains access to the underlying physics and errors, the latter arising from the underlying wave spectra and magnitudes of the RAO around peak events. The framework is considered generalizable and are intended to ultimately enable safer semisubersible operations and reduce weather-waited downtime in regions such as Australia and West Africa.

\section*{Acknowledgments}
This work was supported by the ARC Industrial Transformation Research Hub for Offshore Floating Facilities (OFFshore Hub, https://offshorehub.edu.au/) which is led by The University of Western Australia, delivered with the University of Western Sydney, and funded by the Australian Research Council, Woodside Energy, Shell Australia, Bureau Veritas and Lloyd’s Register Global Technology (Grant No. IH140100012). The work was also supported by the ARC ITRH for Transforming energy Infrastructure through Digital Engineering (TIDE, https://tide.edu.au) which is led by The University of Western Australia, delivered with the University of Wollongong, and supported by the partners listed above with the addition of INPEX Operations Australia, Fugro Australia Marine, Wood Group Kenny Australia, and RPS Group (Grant No. IH200100009). 




\printcredits

\bibliographystyle{cas-model2-names}

\bibliography{semi_heave_forecast}

@article{Brotzge2023,
    author = {Brotzge, J. and Berchoff, D. and Carlis, D. and Carr, F. and Carr, R. and Gerth, J. and Gross, B. and Hamill, T. and Haupt, S. and Jacobs, N. and McGovern, A. and Stensrud, D. and Szatkowski, G. and Szunyogh, I. and Wang, X.},
    title = {Challenges and opportunities in numerical weather prediction},
    journal = {Bulletin of the American Meteorological Society},
    volume = {104(3)},
    year = {2023},
    pages ={E698-E705},
}

@techreport{ECMWF2018,
  author = {R Owens and Tim Hewson},
  title = {{ECMWF} Forecast User Guide},
  year = {2018},
  month = {05/2018},
  publisher = {ECMWF},
  institution = {European Centre for Medium Range Weather Forecasting},
  address = {Reading},
  doi = {10.21957/m1cs7h},
}

@article{Booij1999,
    author = {Booij, N., R.C. Ris and L.H. Holthuijsen},
    title = {A third-generation wave model for coastal regions, Part I, Model description and validation},
    journal = {J. Geophys. Res. C4},
    volume = {104},
    year = {1999},
    pages = {7649-7666},
}

@techreport{Tolman2014,
    author = {Tolman, H.L},
    title = {User manual and system documentation of {WAVEWATCH III} version 4.18. Technical Note 316},
    institution = {NOAA / NWS / NCEP / MMAB},
    year = {2014}, 
}

@article{Wamdi1988,
      author = "{The Wamdi  Group}",
      title = "The {WAM} Model—A Third Generation Ocean Wave Prediction Model",
      journal = "Journal of Physical Oceanography",
      year = "1988",
      publisher = "American Meteorological Society",
      address = "Boston MA, USA",
      volume = "18",
      number = "12",
      pages ="1775 - 1810",
}

@inproceedings{Milne2024,
  title        = {Heave response spectra for a semisubmersible in long period swell},
  author       = {Milne, I. A. and Astfalck, L. C.},
  year         = 2024,
  month        = {December},
  booktitle    = {Proceedings of the 24th Australasian Fluid Mechanics Conference},
  address      = {Canberra, Australia},
  number       = {AFMC2024-221},
}

@inproceedings{Milne2025,
  title        = {Response-Based Forecasting of Vessel Motion: From Physics-Based to Data-Driven Methods},
  author       = {Milne, I. A. and Astfalck, L. C. and Zed, M. and ; Lee-Kopij, J.},
  year         = 2025,
  month        = {May},
  booktitle    = {Proceedings of the Offshore Technology Conference},
  address      = {Houston, Texas, USA},
  number       = {OTC-35614-MS},
}

@report{Zeiger2021,
author = {Zieger, S. and  Greenslade, D. J. M.},
title = {A multiple-resolution global wave model – {AUSWAVE-G3}. {Bureau Research Report - 051.}},
publisher = {Bureau of Meterology},
year = {2021},
}

@article{salvatier2016probabilistic,
  title={Probabilistic programming in Python using {PyMC3}},
  author={Salvatier, John and Wiecki, Thomas V and Fonnesbeck, Christopher},
  journal={PeerJ Computer Science},
  volume={2},
  pages={e55},
  year={2016},
  publisher={PeerJ Inc.}
}

@incollection{BERNARDO2011263,
title = {Modern Bayesian Inference: Foundations and Objective Methods},
editor = {Prasanta S. Bandyopadhyay and Malcolm R. Forster},
booktitle = {Philosophy of Statistics},
publisher = {North-Holland},
address = {Amsterdam},
pages = {263-306},
year = {2011},
volume = {7},
series = {Handbook of the Philosophy of Science},
issn = {18789846},
doi = {https://doi.org/10.1016/B978-0-444-51862-0.50008-3},
author = {José M. Bernardo}
}

@article{gneiting2007probabilistic,
  title={Probabilistic forecasts, calibration and sharpness},
  author={Gneiting, Tilmann and Balabdaoui, Fadoua and Raftery, Adrian E},
  journal={Journal of the Royal Statistical Society Series B: Statistical Methodology},
  volume={69},
  number={2},
  pages={243--268},
  year={2007},
  publisher={Oxford University Press}
}

@article{leung2021forecast,
  title={Forecast verification: Relating deterministic and probabilistic metrics},
  author={Leung, Tsz Yan and Leutbecher, Martin and Reich, Sebastian and Shepherd, Theodore G},
  journal={Quarterly Journal of the Royal Meteorological Society},
  volume={147},
  number={739},
  pages={3124--3134},
  year={2021},
  publisher={Wiley Online Library}
}

@article{sarkar2019spatiotemporal,
  title={Spatiotemporal prediction of tidal currents using {Gaussian} processes},
  author={Sarkar, Dripta and Osborne, Michael A and Adcock, Thomas AA},
  journal={Journal of Geophysical Research: Oceans},
  volume={124},
  number={4},
  pages={2697--2715},
  year={2019},
  publisher={Wiley Online Library}
}

@article{sarkar2018prediction,
  title={Prediction of tidal currents using {Bayesian} machine learning},
  author={Sarkar, Dripta and Osborne, Michael A and Adcock, Thomas AA},
  journal={Ocean Engineering},
  volume={158},
  pages={221--231},
  year={2018},
  publisher={Elsevier}
}

@article{astfalck2019emulation,
  title={Emulation of vessel motion simulators for computationally efficient uncertainty quantification},
  author={Astfalck, LC and Cripps, EJ and Gosling, JP and Milne, IA},
  journal={Ocean Engineering},
  volume={172},
  pages={726--736},
  year={2019},
  publisher={Elsevier}
}

@article {manderson2019,
      author = "A. Manderson and M. D. Rayson and E. Cripps and M. Girolami and J. P. Gosling and M. Hodkiewicz and G. N. Ivey and N. L. Jones",
      title = "Uncertainty Quantification of Density and Stratification Estimates with Implications for Predicting Ocean Dynamics",
      journal = "Journal of Atmospheric and Oceanic Technology",
      year = "2019",
      publisher = "American Meteorological Society",
      address = "Boston MA, USA",
      volume = "36",
      number = "7",
      doi = "10.1175/JTECH-D-18-0200.1",
      pages=      "1313 - 1330"
}

@book{gelman2013bayesian,
  title={{Bayesian data analysis (3rd edition)}},
  author={Gelman, Andrew and Carlin, John B and Stern, Hal S and  Dunson, David B and Vehtari, Aki and Rubin, Donald B},
  year={2013},
  publisher={Chapman and Hall/CRC}
}

@book{bernardo2009bayesian,
  title={{Bayesian theory}},
  author={Bernardo, Jos{\'e} M and Smith, Adrian FM},
  year={2009},
  publisher={John Wiley \& Sons}
}

@article{Astfalck2023, 
title={Evaluating probabilistic forecasts for maritime engineering operations}, 
volume={4}, 
DOI={10.1017/dce.2023.11}, 
journal={Data-Centric Engineering}, publisher={Cambridge University Press}, 
author={Astfalck, Lachlan and Bertolacci, Michael and Cripps, Edward}, 
year={2023}, 
pages={e15}}

@article{Loake2022,
title = {Modelling sea surface wind measurements on Australia’s North-West Shelf},
journal = {Ocean Engineering},
volume = {244},
pages = {110308},
year = {2022},
issn = {0029-8018},
doi = {https://doi.org/10.1016/j.oceaneng.2021.110308},
author = {M.C. Anderson Loake and L.C. Astfalck and E.J. Cripps},
}

@article{MasSoler2018,
title = {Estimating on-site wave spectra from the motions of a semi-submersible platform: An assessment based on model scale results},
journal = {Ocean Engineering},
volume = {153},
pages = {154-172},
year = {2018},
author = {J. Mas-Soler and Alexandre N. Simos and Eduardo A. Tannuri},
}

@Article{Milne2018,
  author   = {I.A. Milne and M. Zed},
  journal  = {Ocean Engineering},
  title    = {Full-scale validation of the hydrodynamic motions of a ship derived from a numerical hindcast},
  year     = {2018},
  issn     = {0029-8018},
  pages    = {83-94},
  volume   = {168},
}

@article{Milne2016,
title = {Validation of a predictive tool for the heading of turret-moored vessels},
journal = {Ocean Engineering},
volume = {128},
pages = {22-40},
year = {2016},
author = {I.A. Milne and S. Delaux and P. McComb},
}

@Misc{HSE2005,
  author = {{Health and Safety Executive}},
  note   = {Research Report 347},
  title  = {Review of the role of response forecasting in decisionmaking for weather-sensitive offshore operations},
  year   = {2005},
}

@Book{Clauss1992,
  author = {G. Clauss and E. Lehmann and C. {\"O}stergaard},
  title  = {Offshore Structures: Volume I: Conceptual Design and Hydromechanics},
  year   = {1992},
}

@Book{Faltinsen1990,
  author = {O.M. Faltinsen},
  title  = {Sea loads on ships and offshore structures},
  publisher = {Cambridge University Press},
  year   = {1990},
}

@article{Ostergaard1987,
title = {Comparison of experimental and theoretical wave actions on floating and compliant offshore structures},
journal = {Applied Ocean Research},
volume = {9},
number = {4},
pages = {192-213},
year = {1987},
author = {C. {\"O}stergaard and T.E. Schellin}
}

@article{Astfalck2018,
title = {Expert elicitation of directional metocean parameters},
journal = {Ocean Engineering},
volume = {161},
pages = {268-276},
year = {2018},
issn = {0029-8018},
doi = {https://doi.org/10.1016/j.oceaneng.2018.04.047},
author = {L.C. Astfalck and E.J. Cripps and J.P. Gosling and M.R. Hodkiewicz and I.A. Milne}
}

@Article{Astfalck2019,
  author   = {L.C. Astfalck and E.J. Cripps and M.R. Hodkiewicz and I.A. Milne},
  journal  = {Ocean Engineering},
  title    = {A {Bayesian} approach to the quantification of extremal responses in simulated dynamic structures},
  year     = {2019},
  issn     = {0029-8018},
  pages    = {594-607},
  volume   = {182},
}

\bio{}
\endbio







\end{document}